\documentclass[12pt]{article}

\def\hybrid{\topmargin -20pt    \oddsidemargin 0pt
        \headheight 0pt \headsep 0pt
        \textwidth 6.25in       
        \textheight 9.5in       
        \marginparwidth .875in
        \parskip 5pt plus 1pt   \jot = 1.5ex}

\hybrid

\catcode`\@=11

\def\marginnote#1{}
\newcount\hour
\newcount\minute
\newtoks\amorpm
\hour=\time\divide\hour by60
\minute=\time{\multiply\hour by60 \global\advance\minute by-\hour}
\edef\standardtime{{\ifnum\hour<12 \global\amorpm={am}%
        \else\global\amorpm={pm}\advance\hour by-12 \fi
        \ifnum\hour=0 \hour=12 \fi
        \number\hour:\ifnum\minute<10 0\fi\number\minute\the\amorpm}}
\edef\militarytime{\number\hour:\ifnum\minute<10 0\fi\number\minute}

\def\draftlabel#1{{\@bsphack\if@filesw {\let\thepage\relax
   \xdef\@gtempa{\write\@auxout{\string
      \newlabel{#1}{{\@currentlabel}{\thepage}}}}}\@gtempa
   \if@nobreak \ifvmode\nobreak\fi\fi\fi\@esphack}
        \gdef\@eqnlabel{#1}}
\def\@eqnlabel{}
\def\@vacuum{}
\def\draftmarginnote#1{\marginpar{\raggedright\scriptsize\tt#1}}

\def\draft{\oddsidemargin -.5truein
        \def\@oddfoot{\sl preliminary draft \hfil
        \rm\thepage\hfil\sl\today\quad\militarytime}
        \let\@evenfoot\@oddfoot \overfullrule 3pt
        \let\label=\draftlabel
        \let\marginnote=\draftmarginnote
   \def\@eqnnum{(\theequation)\rlap{\kern\marginparsep\tt\@eqnlabel}%
\global\let\@eqnlabel\@vacuum}  }

\def\preprint{\twocolumn\sloppy\flushbottom\parindent 2em
        \leftmargini 2em\leftmarginv .5em\leftmarginvi .5em
        \oddsidemargin -.5in    \evensidemargin -.5in
        \columnsep .4in \footheight 0pt
        \textwidth 10.in        \topmargin  -.4in
        \headheight 12pt \topskip .4in
        \textheight 6.9in \footskip 0pt
        \def\@oddhead{\thepage\hfil\addtocounter{page}{1}\thepage}
        \let\@evenhead\@oddhead \def\@oddfoot{} \def\@evenfoot{} }

\def\numberbysection{\@addtoreset{equation}{section}
        \def\theequation{\thesection.\arabic{equation}}}

\def\underline#1{\relax\ifmmode\@@underline#1\else
        $\@@underline{\hbox{#1}}$\relax\fi}

\def\titlepage{\@restonecolfalse\if@twocolumn\@restonecoltrue\onecolumn
     \else \newpage \fi \thispagestyle{empty}\c@page\z@
        \def\thefootnote{\fnsymbol{footnote}} }

\def\endtitlepage{\if@restonecol\twocolumn \else \newpage \fi
        \def\thefootnote{\arabic{footnote}}
        \setcounter{footnote}{0}}  

\catcode`@=12
\relax

\makeatletter
\newcounter{pubctr}
\def\publist{\@ifnextchar[{\@publist}{\@@publist}}
\def\@publist[#1]{\list
        {[\arabic{pubctr}]\hfill}{\settowidth\labelwidth{[999]}
        \leftmargin\labelwidth
        \advance\leftmargin\labelsep
        \@nmbrlisttrue\def\@listctr{pubctr}
        \setcounter{pubctr}{#1}\addtocounter{pubctr}{-1}}}
\def\@@publist{\list
        {[\arabic{pubctr}]\hfill}{\settowidth\labelwidth{[999]}
        \leftmargin\labelwidth
        \advance\leftmargin\labelsep
        \@nmbrlisttrue\def\@listctr{pubctr}}}
 \relax
\makeatother
\newskip\humongous \humongous=0pt plus 1000pt minus 1000pt

\newif\ifdtup

\relax

\def\d{\partial}

\def\sqr#1#2{{\vcenter{\vbox{\hrule height.#2pt\hbox{\vrule width.#2pt 

height#1pt \kern#1pt \vrule width.#2pt}\hrule height.#2pt}}}}

\def\=d{\,{\buildrel\rm def\over =}\,}

\def\i3p{\p32\int d^3p}

\def\As{A\hbox to 1pt{\hss /}}
\def\np4{\int d^4p_1\cdots d^4p_{n-1}\, }

\def\Tr{{\rm Tr}\, }

\def\nx4{\int d^4x_1\ldots d^4x_n\, }

\def\kon#1#2{\vbox{\halign{##&&##\cr
\lower4pt\hbox{$\scriptscriptstyle\vert$}\hrulefill &
\hrulefill\lower4pt\hbox{$\scriptscriptstyle\vert$}\cr $#1$&
$#2$\cr}}}

\def\konv#1#2#3{\hbox{\vrule height12pt depth-1pt}
\vbox{\hrule height12pt width#1cm depth-11.6pt}
\hbox{\vrule height6.5pt depth-0.5pt}
\vbox{\hrule height11pt width#2cm depth-10.6pt\kern5pt
      \hrule height6.5pt width#2cm depth-6.1pt}
\hbox{\vrule height12pt depth-1pt}
\vbox{\hrule height6.5pt width#3cm depth-6.1pt}
\hbox{\vrule height6.5pt depth-0.5pt}}
\def\konu#1#2#3{\hbox{\vrule height12pt depth-1pt}
\vbox{\hrule height1pt width#1cm depth-0.6pt}
\hbox{\vrule height12pt depth-6.5pt}
\vbox{\hrule height6pt width#2cm depth-5.6pt\kern5pt
      \hrule height1pt width#2cm depth-0.6pt}
\hbox{\vrule height12pt depth-6.5pt}
\vbox{\hrule height1pt width#3cm depth-0.6pt}
\hbox{\vrule height12pt depth-1pt}}

\def\konw#1#2#3{\hbox{\vrule height12pt depth-1pt}
\vbox{\hrule height12pt width#1cm depth-11.6pt}
\hbox{\vrule height6.5pt depth-0.5pt}
\vbox{\hrule height12pt width#2cm depth-11.6pt \kern5pt
      \hrule height6.5pt width#2cm depth-6.1pt}
\hbox{\vrule height6.5pt depth-0.5pt}
\vbox{\hrule height12pt width#3cm depth-11.6pt}
\hbox{\vrule height12pt depth-1pt}}

\def\i{{\rm int}}

\def\r{{\rm ret}}
\def\a{{\rm av}}

\def\m3{{\mu_1\mu_2\mu_3}}

\def\p{{(+)}}

\def\be{\begin{equation}}       \def\eq{\begin{equation}}
\def\ee{\end{equation}}         \def\eqe{\end{equation}}

\def\bea{\begin{eqnarray}}      \def\eqa{\begin{eqnarray}}
\def\ena{\end{eqnarray}}        \def\eea{\end{eqnarray}}
                                \def\eqae{\end{eqnarray}}

\def\ba{\begin{array}}
\def\ea{\end{array}}
\def\unit{1 \hskip-.3em \raise2pt\hbox{$ \scriptstyle |$ } }

\def\a{\alpha}

\def\d{\delta}
\def\g{\gamma}

\def\i{\iota}

\def\m{\mu}

\def\p{\pi}                
\def\r{\rho}                                     
\def\s{\sigma}                                   

\def\G{\Gamma}

\def\cl{{\cal L}}

\def\cn{{\cal N}}

\def\bop#1{\setbox0=\hbox{$#1M$}\mkern1.5mu
        \vbox{\hrule height0pt depth.04\ht0
        \hbox{\vrule width.04\ht0 height.9\ht0 \kern.9\ht0
        \vrule width.04\ht0}\hrule height.04\ht0}\mkern1.5mu}
\def\Box{{\mathpalette\bop{}}}                        
\def\pa{\partial}                              

\def\>{\rangle} 

\def\<{\langle} 
\def\Dsl{D \hskip-.6em \raise1pt\hbox{$ / $ } }

\def\to{\rightarrow}

\def\1ov4{{1\over 4}}

\def\Tr{{\rm Tr}}

\def\pa{\partial}

\def\pa{\partial}

\def\nonu{\nonumber \\{}}

\def\IR{\relax{\rm  I\kern-.18em R}}
\def\inv{^{\raise.15ex\hbox{${\scriptscriptstyle -}$}\kern-.05em 1}}

\def\be{\begin{equation}}
\def\ee{\end{equation}}
\def\bea{\begin{eqnarray}}
\def\eea{\end{eqnarray}}
\def\g0{g_{(0)}}
\def\gi{g_{(0)}^{-1}}
\def\Tr{{\rm Tr}}

\begin{document}
\thispagestyle{empty} \setcounter{footnote}{0}
\begin{flushright}
CERN-TH/98-188\\
\vskip -.1 cm
KUL-TF-98/21\\

\vskip -.1 cm
hep-th/9806087\\
\end{flushright}

\vskip 1,4cm
\begin{center}
{\Large \bf The Holographic Weyl Anomaly}

  \vskip 1.8cm

{\bf M. Henningson}
\vskip 0,2cm
{\it Theory Division, CERN \\
CH-1211 Geneva 23, Switzerland\\
{\tt henning@nxth04.cern.ch }}
  \\
  \vskip .9cm
  {\bf K. Skenderis}
  \vskip 0,2cm
  {\it Instituut voor Theoretische Fysica, KU Leuven\\
    Celestijnenlaan 200D, B-3001 Leuven, Belgium\\
    {\tt kostas.skenderis@fys.kuleuven.ac.be}}

\end{center}
\vskip 1 cm

\centerline {\bf Abstract}
We calculate the Weyl anomaly for conformal field theories that can
be described via the adS/CFT correspondence. This entails regularizing
the gravitational part of the corresponding supergravity action in
a manner consistent with general covariance. Up to a constant, the
anomaly only depends on the dimension $d$ of the manifold on which
the conformal field theory is defined. We present concrete
expressions for the anomaly in the physically relevant cases $d=2, 4$ and $6$.
In $d=2$ we find for the central charge $c=3 l/ 2 G_N$, in agreement with
considerations based on the asymptotic symmetry algebra of $adS_3$.
In $d=4$ the anomaly agrees precisely with that of the corresponding
$\cn=4$ superconformal $SU(N)$ gauge theory. The result in $d=6$ 
provides new information for the $(0, 2)$ theory, since its 
Weyl anomaly has not been computed previously. The anomaly
in this case grows as $N^3$, where $N$ is the number of coincident
$M5$ branes, and it vanishes for a Ricci-flat background.

\vskip 0,2cm
\noindent

\vfill

\begin{flushleft}
CERN-TH/98-188\\
\vskip -.1 cm
KUL-TF-98/21\\
\vskip -.1 cm
June 1998
\end{flushleft}

\newpage

\section{Introduction}

At low energies, the worldvolume theory on $N$ coincident $p$-branes
in $M$-theory or string theory decouples from the bulk theory and
can be studied on its own. In some cases, the worldvolume theory
constitutes a conformal field theory (CFT). This is true, for example,
for $D3$-branes in type $IIB$ string theory and for
five-branes in $M$-theory, which give rise
to the $d=4$ ${\cal N}=4$ superconformal $SU(N)$ gauge theory and a
$d=6$ $(0, 2)$ superconformal field theory, respectively.
It has recently been conjectured by Maldacena \cite{malda},
following earlier work on black holes \cite{kleb1}--\cite{KSKS},
that these conformal field theories are dual to $M$-theory or string
theory in the background describing the near-horizon brane configuration.
This equivalence may also be inferred by observing that  
the brane configuration can be mapped to its near-horizon limit \cite{BPS} by
means of certain duality transformations \cite{hyun}. However, this argument
is as yet incomplete, since these duality transformations are not fully
understood, as they involve the time coordinate. The correspondence
between string theory in a specific background and conformal field
theories is a realization of the holographic principle advocated by
't Hooft \cite{tHooft} and Susskind \cite{Susskind} in that it
describes a $(d+1)$-dimensional theory containing gravity in terms of degrees
of freedom on a $d$-dimensional hypersurface.

The conjectured correspondence was clarified by Gubser, Klebanov and
Polyakov \cite{GKP} and by Witten \cite{Witten}
as follows: the supergravity background is a
product of a compact manifold and a $(d + 1)$-dimensional manifold
$X_{d + 1}$ with a boundary (``horizon'') $M_d$. The conformal field
theory is defined on $M_d$. There is a one-to-one relationship
between operators ${\cal O}$ of the conformal field
theory and the fields $\phi$ of the supergravity theory.
In particular, gauge fields in the bulk couple to global
currents in the boundary.
The presence of a boundary means that the supergravity action
functional $S[\phi]$ must be supplemented by a boundary condition
for $\phi$ parametrized by a field $\phi^{(0)}$ on $M_d$.
The partition function is then a functional of the boundary
conditions
\be
Z_{string}[\phi^{(0)}] = \int_{\phi^{(0)}} {\cal D} \phi \exp \left(-
S[\phi]\right) ,
\ee
where the subscript $\phi^{(0)}$ on the integral sign indicates that
the functional integral is over field configurations $\phi$ that
satisfy the boundary condition given by $\phi^{(0)}$. The conjecture
states that the string (or $M$-theory) partition function,
as a functional of $\phi^{(0)}$, equals the generating functional of
correlation functions in the conformal field theory:
\be \label{CFT}
Z_{CFT}[\phi^{(0)}] = \left\langle \exp \int_{M_d} d^d x {\cal O}
\phi^{(0)} \right\rangle .
\ee
The fields $\phi^{(0)}$ act as sources for the operators of the
conformal field theory. Notice that the
bulk theory only sees, through the boundary values of its fields,
the abstract conformal field theory
and not the elementary fields that may realize it.
The partition function (\ref{CFT}) may also be viewed as
describing the coupling of conformal matter to
conformal supergravity \cite{Liu-Tseytlin}. The sources
$\phi^{(0)}$ constitute conformal supermultiplets.

The relationship just described is conjectured to hold for any number
$N$ of coincident branes. However, in most cases one can reliably
compute
the string partition function only for large $N$. The reason is that
the backgrounds involve RR forms whose coupling to perturbative
strings
is through $D$-branes. Therefore a complete string calculation
is rather difficult to perform. However, if the number of branes is
large, the characteristic length scale of the
supergravity background is large compared to the string scale
(or the Planck scale in the case of $M$-theory), and one can trust the
supergravity approximation. In addition the string coupling
may be chosen small. Under these circumstances,
the string partition function reduces to the exponential
of the supergravity action functional evaluated for a field
configuration $\phi^{cl}(\phi^{(0)})$ that solves the classical
equations of motion and satisfies the boundary conditions given by
$\phi^{(0)}$
\be
Z^{tree}_{string}[\phi^{(0)}] = \exp \left(-S[\phi^{cl} (\phi^{(0)})]
\right) .
\ee

An operator of particular importance in any conformal field theory is
the energy-momentum tensor $T_{ij}$, $i, j = 1, \ldots, d$. The
corresponding bulk gauge field is the metric
$\hat{G}_{\mu \nu}$, $\mu, \nu = 0, \ldots, d$ on $X_{d + 1}$.
In the supergravity backgrounds under consideration, the metric
$\hat{G}_{\mu \nu}$ does not induce a
unique metric $\g0$ on the boundary $M_d$, because it has a 
second-order pole there. However it does determine a conformal equivalence
class or conformal structure $[\g0]$ of metrics on $M_d$. To get a
representative $\g0$, we pick a function $\rho$ on $X_{d + 1}$ with a
simple zero on $M_d$ and restrict $\rho^2 \hat{G}_{\mu \nu}$ to $M_d$.
Different choices of the function $\rho$ yield different metrics on
$M_d$ in the same conformal equivalence class. The field that
specifies the boundary condition of the metric is thus a conformal
structure $[\g0]$. This means that, at least naively, the trace
of the energy-momentum tensor decouples.

In this paper we wish to determine the dependence of 
the boundary theory partition function
(or zero-point function) on a given representative $\g0$ of the
conformal
structure. In other words, we shall study whether the trace of the
energy-momentum decouples. Since we examine correlation functions
that only involve the energy-momentum tensor, the only relevant part
of the bulk action is the gravitational one. Therefore we set 
all other fields to zero. At tree level, we then need to solve the classical
supergravity equations of motion on $X_{d + 1}$ subject to the conditions that
the metric $\hat{G}_{\mu \nu}$ on $X_{d + 1}$ induces a given
conformal structure $[\g0]$ on $M_d$ and all other fields vanish there.
In the theories under consideration,
this means that $\hat{G}_{\mu \nu}$ fulfils Einstein's equations
\be\label{Einstein}
\hat{R}_{\mu \nu} - \frac{1}{2} \hat{G}_{\mu \nu} \hat{R} =
\Lambda \hat{G}_{\mu \nu},
\ee
with some cosmological constant $\Lambda$ and that all other fields
vanish identically on $X_{d + 1}$. According to a theorem due to
Graham and Lee \cite{Graham-Lee}, up to
diffeomorphism, there is a unique such metric $\hat{G}_{\mu \nu}$. 
(Actually the theorem has been proved for the case when $X_{d + 1}$ is
topologically a ball $B_{d + 1}$ so that $M_d$ is a sphere $S^d$ and
the conformal structure $[\g0]$ on $M_d$ is sufficiently close to the
standard (conformally flat) one.) The conformal field theory
effective action (strictly speaking, the generating functional of the
connected graphs) $W_{CFT}[\g0] = - \log Z_{CFT}[\g0]$ is then
given by evaluating the action functional
\bea\label{action}
S[\hat{G}_{\mu \nu}] & = & S_{bulk} + S_{boundary} \cr
& = & \frac{1}{16 \pi G^{(d + 1)}_N} \left[\int_{X_{d + 1}} d^{d + 1}
x
\sqrt{\det \hat{G}} \left(\hat{R} + 2 \Lambda \right)
+\int_{M_d} d^d x \sqrt{\det
\tilde{g}} \left( 2 D_{\mu} n^\mu + \alpha \right)\right]
\eea
for this metric. Here $\tilde{g}$ is the metric induced on $M_d$
from $\hat{G}$, and $n^\mu$ is a unit normal vector to $M_d$. The
bulk term is of course the usual Einstein-Hilbert action with a
cosmological constant. The inclusion of the first boundary term is
necessary on a manifold with boundary in order to get an action that
depends only on first derivatives of the metric \cite{Gibbons-Hawking}.
The possibility of including the second boundary term with some
coefficient $\alpha$ was first discussed in \cite{Liu-Tseytlin}.

The above description might seem to indicate that the conformal field
theory effective action $W_{CFT}[\g0]$ only depends on the
conformal equivalence class of the metric on $M_d$. This is of course
as it should be in a truly conformally invariant theory. However, the
action functional (\ref{action}) does not make sense for the metric
$\hat{G}_{\mu \nu}$ determined by (\ref{Einstein}) and the boundary
conditions. Indeed, the bulk term of the action diverges because of
the infinite volume of $X_{d + 1}$. The boundary terms are also
ill-defined, since the induced metric $\tilde{g}_{ij}$ on $M_d$
diverges because of the double pole of $\hat{G}_{\mu \nu}$. The
action should therefore be regularized in a way that preserves general
covariance, so that the divergences can be cancelled by the addition
of local counterterms. As we will see shortly, this regularization
entails picking a particular, but arbitrary, representative $\g0$ of
the conformal structure $[\g0]$ on $M_d$. In this way, one obtains a
finite effective action, which, however, will depend on the choice of this
representative metric. Conformal invariance is thus explicitly broken
by a so-called conformal or Weyl anomaly. The anomaly, which is usually
perceived as a UV effect, thus arises from an IR-divergence in the
bulk theory. This is an example of a more general IR-UV connection
that applies to holographic theories \cite{Susskind-Witten}.

In this paper, we will calculate the Weyl anomaly for conformal 
field theories that can be derived from a supergravity theory,
 as described above. In the next
section, we will describe the regularization procedure and the
computation of the anomaly in general. In the last section,  we
evaluate the anomaly in the physically relevant cases $d = 2, 4, 6$.
For $d = 2$ and $d = 4$ we compare with the known anomaly for the
$adS_3$ boundary conformal field theory and the $d = 4$ ${\cal N} =
4$ superconformal $SU(N)$ gauge theory respectively, and find perfect
agreement. For $d = 6$ there is no corresponding calculation of the 
Weyl anomaly, so our result provides new information about the
$(0, 2)$ superconformal field theory.

\section{The regularization procedure}
A regularization scheme that preserves general covariance was
described
in \cite{Witten}. As discussed above, up to
diffeomorphisms, there is a  unique Einstein metric $\hat{G}$ 
on $X_{d + 1}$ that induces a given a conformal structure $[\g0]$ 
on the boundary $M_d$.
We now pick a metric $\g0$ on $M_d$ in the given conformal equivalence
class. According to a theorem due to Fefferman and Graham
\cite{Fefferman-Graham}, there is a distinguished coordinate
system $(\rho, x^i)$ on $X_{d + 1}$ in which $\hat{G}$ takes the form
\be\label{metric}
\hat{G}_{\mu \nu} d x^\mu d x^\nu = \frac{l^2}{4} \rho^{-2} d \rho d
\rho + \rho^{-1} g_{ij} d x^i d x^j ,
\ee
where the tensor $g$ has the limit $\g0$ as one approaches the
boundary
represented by $\rho = 0$. The length scale $l$ is related to the
cosmological constant $\Lambda$ as $\Lambda = - \frac{d (d - 1)}{2
l^2}$. Einstein's equations for $\hat{G}$ amount to
\bea
\rho \left(2 g^{\prime\prime} - 2 g^\prime g^{-1} g^\prime + \Tr
(g^{-1} g^\prime) g^\prime \right) + l^2 {\rm Ric} (g) - (d - 2)
g^\prime - \Tr (g^{-1} g^\prime) g & = & 0 \cr
(g^{-1})^{jk} \left(\nabla_i g_{jk}^\prime - \nabla_k g_{ij}^\prime
\right) & = & 0 \cr
\Tr (g^{-1} g^{\prime\prime}) - \frac{1}{2} \Tr (g^{-1} g^\prime
g^{-1}
g^\prime) & = & 0 , \label{eqn}
\eea
where differentiation with respect to $\rho$ is denoted with a prime,
$\nabla_i$ is the covariant derivative constructed from the metric
$g$
and ${\rm Ric} (g)$ is the Ricci tensor\footnote{
Our conventions are as follows
$R_{ijk}{}^l=\pa_i \G_{jk}{}^l + \G_{ip}{}^l \G_{jk}{}^p - i
\leftrightarrow j$ and $R_{ij}=R_{ikj}{}^k$.} of $g$.

In the case when $d$ is odd, these equations can be solved order by
order in $\rho$ so that
\be\label{odd}
g = \g0 + \rho g_{(2)} + \rho^2 g_{(4)} + \ldots ,
\ee
where the tensor $g_{(k)}$ is given by some covariant expression in
the
boundary metric $\g0$, its Riemann tensor and the corresponding
covariant derivative. Throughout this paper, a subscript in
parentheses
on a quantity indicates the number of derivatives with respect to
$x^i$. In the case when $d$ is even, this procedure breaks down at
order $d/2$ in $\rho$, where a logarithmic term appears:
\be\label{even}
g = \g0 + \rho g_{(2)} + \ldots + \rho^{d/2} g_{(d)} + \rho^{d/2}
\log \rho \ h_{(d)} + {\cal O} (\rho^{d/2 + 1}) .
\ee
The tensors $g_{(k)}$ for $k = 0, 2, \ldots, d - 2$ are again
covariant. The same is true for $\Tr (\gi g_{(d)})$ but not for the
complete tensor $g_{(d)}$. Finally, $\Tr (\gi h_{(d)})$ vanishes
identically.

The regularization procedure now amounts to restricting
the bulk integral to the domain $\rho > \epsilon$ for some cutoff
$\epsilon > 0$ and evaluating the boundary integrals at $\rho =
\epsilon$. The regulated action evaluated for the metric $\hat{G}$ is
thus $(16 \pi G^{(d + 1)}_N)^{-1} \int d^d x \, {\cal L}$, where
\bea\label{Lagrangian}
{\cal L} & = & \frac{d}{l} \int_\epsilon d \rho \rho^{-d/2 - 1}
\sqrt{\det g} \cr
& & + \left. \rho^{-d/2} \left( - \frac{2 d}{l} \sqrt{\det g}  +
\frac{4}{l} \rho \partial_\rho {\sqrt{ \det g}} + \alpha \sqrt{\det
g}
\right) \right|_{\rho = \epsilon} .
\eea
In the first term, which arises from the bulk part of the action,  we
have used the fact that $\hat{G}$ is an Einstein metric so that $\hat{R} + 2
\Lambda = - \frac{4}{d - 1} \Lambda = \frac{2 d}{l^2}$.

For $d$ odd, it follows from (\ref{odd}) that $\sqrt{\det g}$ is a
power series in $\rho$ with covariant coefficients. For $d$ even,
this is true up to and including the $\rho^{d/2}$ terms. (The higher-order
non-covariant corrections will play no role in the sequel). The
Lagrangian (\ref{Lagrangian}) can therefore be written as
\be\label{Lodd}
{\cal L} = \sqrt{\det \g0} \left( \epsilon^{-d/2} a_{(0)} +
\epsilon^{-d/2 + 1} a_{(2)} + \ldots + \epsilon^{-1/2} a_{(d - 1)}
\right) + {\cal L}_{fin}
\ee
for $d$ odd, and as
\be\label{Leven}
{\cal L} = \sqrt{\det \g0} \left( \epsilon^{-d/2} a_{(0)} +
\epsilon^{-d/2 + 1} a_{(2)} + \ldots + \epsilon^{-1} a_{(d - 2)}
- \log \epsilon \, a_{(d)} \right) + {\cal L}_{fin} ,
\ee
for $d$ even, where ${\cal L}_{fin}$ is finite in the $\epsilon \to 0$ 
limit. All the $a_{(k)}$ coefficients are covariant, so the
divergent terms can be cancelled by subtracting covariant
counterterms, as promised. The logarithmic divergence 
that appears for $d$ even comes only from the bulk
integral.

After subtraction of the divergent counterterms, we are left with a
renormalized effective action 
$(16 \pi G^{(d + 1)}_N)^{-1} \int d^d x\, {\cal L}_{fin}$ 
with a finite limit as $\epsilon$ goes to zero. Its
variation under a conformal transformation $\delta \g0 = 2 \delta
\sigma \g0$ for an infinitesimal parameter function $\delta \sigma$
is
of the form
\be
\d \cl_{fin}=-\int_{M_d} d^d x \sqrt{\det \g0} \delta \sigma {\cal A} ,
\ee
and we would like to calculate the anomaly ${\cal A}$. For $d$ odd,
${\cal A}$ in fact vanishes, whereas for $d$ even
\be\label{anomaly}
{\cal A} = \frac{1}{16 \pi G^{(d + 1)}_N} (-2 a_{(d)}) .
\ee
To see this, we note that for a {\it constant} parameter $\delta
\sigma$, the regulated Lagrangian (\ref{Lodd}) or (\ref{Leven}) is
invariant under the combined transformation $\delta \g0 = 2 \delta
\sigma \g0$ and $\delta \epsilon = 2 \delta \sigma \epsilon$. The
terms
proportional to negative powers of $\epsilon$ are separately
invariant,
so the variation of the finite part plus the variation of the
logarithmically divergent term (for $d$ even) must vanish. Since
$\log \epsilon$ transforms with a shift and $\sqrt{\det g_{(0)}} a_{(d)}$ 
itself is invariant, we get (\ref{anomaly}).

On general grounds \cite{Bonora-Pasti-Bregola, Deser-Schwimmer},
the coefficient $a_{(d)}$ that appears in the anomaly (\ref{anomaly})
must be of the form
\be\label{decomposition}
a_{(d)} = d l^{d - 1} \left(E_{(d)} + I_{(d)} + D_i J_{(d - 1)}^i
\right) ,
\ee
where $E_{(d)}$ is proportional to the $d$-dimensional Euler density
and $I_{(d)}$ is a conformal invariant. These terms are referred to as
the type A and the type B anomaly, respectively, in
\cite{Deser-Schwimmer}. The dimension of the space of conformal 
invariants grows with $d$.
The $D_i J_{(d - 1)}^i$ term, where $D_i$ is
the covariant derivative constructed from the boundary metric $\g0$,
is trivial in the sense that it can be cancelled by the variation of a
finite covariant counterterm added to the action. To see this, notice
that a covariant counterterm will be, in particular, scale invariant.
Making the parameter of the scale transformation local amounts to
computing the Noether current for scale transformations.
Thus, the result of the variation is $\d \s D_i J^i$.
However, local scale transformations are just
Weyl transformations. Thus terms of the form $D_i J^i$ can be
obtained by variation of covariant counterterms.

The coefficients of the various independent contributions (properly 
normalized) in (\ref{decomposition}) are closely related to renormalization 
group equations, and they reflect the matter content of 
the superconformal theory. Using Ward identities 
one can relate them to Schwinger terms in the OPEs of the energy-momentum 
tensor \cite{OP}--\cite{AFGJ}. For a recent application, see \cite{kleb4}.
Our results for $d=6$ can be similarly used to determine Schwinger 
terms in the OPEs of the $(0, 2)$ theory.

\section{Evaluation of the anomaly}
In this section, we will perform the above procedure in the physically
relevant cases $d = 2, 4, 6$ and give concrete formulas for the
quantities $E$, $I$ and $J^i$ appearing in (\ref{decomposition}).
As we have mentioned in the previous paragraph, the logarithmic
divergence comes only from the bulk integral. It is completely
straightforward to obtain $a_{(d)}$. One only needs to
expand $\sqrt{\det g}$ up to appropriate order in $\r$.
In the formulae below, we further simplify the result by eliminating
$\Tr(g_{(0)}^{-1} g_{(n)})$, for $n>2$, by using the third equation
in (\ref{eqn}). We raise and lower indices with the boundary metric
$g_{(0)}$ and its  inverse $g_{(0)}^{-1}$.
The Riemann tensor and the covariant derivative
constructed from $g_{(0)}$ are denoted $R^i{}_{jkl}$ and $D_i$,
respectively.

\subsection{$d = 2$ and the asymptotic symmetry algebra of $adS_3$}
Calculating
\be
a_{(2)} = l \, \Tr (\gi g_{(2)})
\ee
and decomposing it according to (\ref{decomposition}), we get
\bea
E_{(2)} & = & \frac{1}{4} R \cr
I_{(2)} & = & 0 \cr
J_{(1)}^i & = & 0 .
\eea
(There is in fact no non-trivial conformal invariant $I$ in this
dimension.) Writing the anomaly in the form
\be
{\cal A} = - \frac{c}{24 \pi} R
\ee
(in our conventions a free boson contributes to the anomaly $-1/24 \p R$)
we thus get
\be
c = \frac{3 l}{2 G^{(3)}_N} .
\ee
This agrees with the value of the conformal anomaly $c$ as computed
in \cite{Brown-Henneaux} by considering the asymptotic symmetry algebra
of
$adS_{3}$.

\subsection{$d = 4$ and ${\cal N} = 4$ super Yang-Mills theory}
In this case one finds
\bea
a_{(4)}&=& l^3 \frac{1}{2} \left([\Tr (\gi g_{(2)})]^2 
- \Tr [(\gi g_{(2)})^2]\right) \cr
&=&l^3 \left(-{1 \over 8} R^{ij} R_{ij} + {1 \over 24} R^2 \right).
\label{d4an}
\eea
Notice that this expression vanishes for a Ricci-flat background.
A check on our calculation is whether (\ref{d4an}) can be 
rewritten in the form (\ref{decomposition}). Indeed, this is possible, 
and we obtain
\bea
E_{(4)} & = &  \frac{1}{64} \left(R^{ijkl} R_{ijkl} - 4 R^{ij}
R_{ij}
+R^2 \right) \cr
I_{(4)} & = & -\frac{1}{64} \left(R^{ijkl} R_{ijkl} - 2 R^{ij} R_{ij}
+
\frac{1}{3} R^2 \right) \cr
J_{(3)}^i & = & 0 .
\eea
(Up to a constant, $I_{(4)}$ is in fact the unique conformal
invariant with four derivatives in this dimension, namely the Weyl tensor
contracted with itself.) We now use the fact that $g^{(5)}_N =
g^{(10)}_N / Vol(S^5)$, where $Vol(S^5)=l^5 \pi^3$ is the volume
of the compactification five-sphere of radius $l$, and
$G^{(10)}_N = 8 \pi^6 g_{\rm str}^2$ is the
ten-dimensional Newton's constant (the $\a'$'s cancel out in the 
Maldacena limit). Furthermore, $l$ is
related to the number $N$ of $D3$-branes as
$l = (4 \pi g_{str} N)^{1/4}$ \cite{malda}.
Putting everything together, we get
\be\label{fouranomaly}
{\cal A} = -\frac{N^2}{\pi^2} \left(E_{(4)} + I_{(4)} \right) .
\ee
This should be compared with the conformal anomaly of the $d = 4$
${\cal N} = 4$ superconformal $SU(N)$ gauge theory. The conformal
anomaly of a theory with $n_s$ scalar fields, $n_f$ Dirac fermions
and
$n_v$ vector fields is \cite{4danom10}
\be \label{anom}
-\frac{1}{90 \pi^2} (n_s + 11 n_f + 62 n_v) E_{(4)} - \frac{1}{30\pi^2}
(n_s + 6 n_f + 12 n_v) I_{(4)} .
\ee
The anomaly of the $\cn=4$ $SU(N)$ super Yang-Mills multiplet
is equal to $N^2-1$ (as all fields are in the adjoint)
times (\ref{anom}) for $n_s = 6$, $n_f = 2$ and $n_v = 1$.
Thus, in the large $N$ limit we obtain exact agreement with 
(\ref{fouranomaly}). This is perhaps surprising since the result
(\ref{anom}) is derived using free fields whereas our result 
is about the full interacting $\cn=4$ $SU(N)$ superconformal
field theory. This indicates that there must be a non-renormalization 
theorem that protects these coefficients. 

Various orbifolding procedures \cite{Kachru-Silverstein, Witten2} change 
the volume of the compactification space and also give rise to other gauge 
groups. It is easy to check that the anomalies still work out correctly.

\subsection{$d = 6$ and tensionless strings}
Following \cite{Bonora-Pasti-Bregola}, we introduce
\bea
(K_1, \ldots, K_{11}) & = & \left(R^3, R R_{ij} R^{ij}, R R_{ijkl}
R^{ijkl}, R_i{}^j R_j{}^k R_k{}^i, R^{ij} R^{kl} R_{iklj}, \right.
\cr
& & R_{ij} R^{iklm} R^j{}_{klm}, R_{ijkl} R^{ijmn} R^{kl}{}_{mn},
R_{ijkl} R^{imnl} R^j{}_{mn}{}^k, \cr
& & \left. R \Box R, R_{ij} \Box R^{ij}, R_{ijkl} \Box R^{ijkl} \right) .
\eea
The six-dimensional Euler density is then proportional to
\be
E_0 = K_1 - 12 K_2 + 3 K_3 + 16 K_4 -24 K_5 - 24 K_6 + 4 K_7 + 8 K_8
\ee
and
\bea
I_1 & = & \frac{19}{800} K_1 - \frac{57}{160} K_2 + \frac{3}{40} K_3
+
\frac{7}{16} K_4 - \frac{9}{8} K_5 - \frac{3}{4} K_6 + K_8 \cr
I_2 & = & \frac{9}{200} K_1 - \frac{27}{40} K_2 + \frac{3}{10} K_3 +
\frac{5}{4} K_4 - \frac{3}{2} K_5 - 3 K_6 + K_7 \cr
I_3 & = & K_1 - 8 K_2 - 2 K_3 + 10 K_4 - 10 K_5 - \frac{1}{2} K_9 + 5
K_{10} - 5 K_{11}
\eea
form a basis for conformal invariants with six derivatives.

We have
\bea
a_{(6)} & = & l^5 \left( {1 \over 8} [\Tr \gi g_{(2)}]^3
-{3 \over 8} \Tr[\gi g_{(2)}] \Tr[(\gi g_{(2)})^2]\right. \nonu
&& \left.+{1 \over 2} \Tr[(\gi g_{(2)})^3]
- \Tr[\gi g_{(2)} \gi g_{(4)}] \right).
\eea
Evaluating this expression we obtain
\bea
a_{(6)} &=& {l^5 \over 64} \left( 
-{1 \over 2} R R^{ij} R_{ij} + {3 \over 50} R^3
+ R^{ij} R^{kl} R_{ikjl} \right.\nonu
&&\left.+ {1 \over 5} R^{ij} D_i D_j R - {1 \over 2} R^{ij} \Box R_{ij} 
+ {1 \over 20} R \Box R \right). \label{d6an}
\eea
Observe that the above expression vanishes in a Ricci-flat background.
The next task is to put this expression in the form 
(\ref{decomposition}). This is a nice check on our calculation.
The result is 
\bea
E_{(6)} & = & \frac{1}{6912} E_0 \cr
I_{(6)} & = & \frac{1}{1152} 
\left(-{ 10 \over 3} I_1 - \frac{1}{6} I_2
+\frac{1}{10} I_3 \right) \cr
J_{(5)}^i & = & - {1 \over 1152} [- R^{ijkl} D^m R_{mjkl}
+2(R_{jk} D^i R^{jk} -R_{jk} D^j R^{ki})] \nonu
&&+{1 \over 720} R^{ij} D_j R +{17 \over 11520} R D^i R
\eea
We now use the fact that $G^{(7)}_N=G^{(11)}_N/Vol(S^4)$, where 
$Vol(S^4)=R_{sph}^4 (8 \p^2/3)$ and $R_{sph}=l_{Planck} (\pi N)^{1/3}$
is the radius of the compactification sphere. In addition, the 
eleven-dimensional Newton's constant is equal to 
$G^{(11)}_N = 16 \pi^7 l_{Planck}^9$, and the characteristic 
length $l$ is $l = 2 l_{Planck} (\pi N)^{1/3}$. 
Putting every together we get for the anomaly
\be
{\cal A} = -\frac{4 N^3}{\pi^3} \left(E_{(6)} + I_{(6)} + D_i J_{(5)}^i
\right) .
\ee
The anomaly for a $(0, 2)$ tensor multiplet has not yet been
calculated. However, we see that the anomaly grows as $N^3$, in
agreement with considerations based on the entropy of the brane
system \cite{Klebanov-Tseytlin, kleb4}. This growth is presumably 
related to the appearance of tensionless strings when multiple
fivebranes coincide.

\section*{Acknowledgements}
We would like to thank each other's institute for hospitality and
financial support during part of this work.

\end{document}